%% file: paper.tex
\documentclass[conference]{IEEEtran}
\IEEEoverridecommandlockouts
\usepackage{cite}
\usepackage{amsmath,amssymb,amsfonts}
\usepackage{algorithmic}
\usepackage{graphicx}
\usepackage{textcomp}
\usepackage{xcolor}
\def\BibTeX{{\rm B\kern-.05em{\sc i\kern-.025em b}\kern-.08em
    T\kern-.1667em\lower.7ex\hbox{E}\kern-.125emX}}

\usepackage[caption=false,font=footnotesize]{subfig}
\usepackage{multirow}
\usepackage{cleveref}
\usepackage{pgfplots}
\pgfplotsset{compat=1.18}
\usepackage{pgfplotstable}

\definecolor{colorSupervised}{RGB}{0,0,0}        % Black
\definecolor{colorSimCLR}{RGB}{230,159,0}        % Orange
\definecolor{colorBYOL}{RGB}{86,180,233}         % Sky Blue
\definecolor{colorBarlow}{RGB}{102,67,57}        % Brown
\definecolor{colorVICReg}{RGB}{213,94,0}         % Vermillion
\definecolor{colorDINO}{RGB}{0,114,178}          % Blue
\definecolor{colorLeJEPA}{RGB}{153,153,153}      % Grey
\definecolor{colorHypersolid}{RGB}{170,51,119}   % Deep magenta
\definecolor{colorReSA}{RGB}{88,24,69}         % YDark plum
\definecolor{colorSwAV}{RGB}{128,128,128}        % Darker gray

\pgfplotscreateplotcyclelist{mycolorlist}{
    {colorSupervised, mark=*, mark size=1.5pt, thick},         % Solid Circle
    {colorSimCLR, mark=square*, mark size=1.5pt},             % Solid Square
    {colorBYOL, mark=triangle*, mark size=1.8pt},             % Solid Triangle
    {colorSwAV, mark=triangle, mark size=1.8pt},
    {colorBarlow, mark=diamond*, mark size=1.8pt},            % Solid Diamond
    {colorVICReg, mark=o, mark size=1.5pt, thick},            % Open Circle
    {colorDINO, mark=square, mark size=1.5pt, thick},         % Open Square
    {colorReSA, mark=star, mark size=1.5pt},
    {colorHypersolid, mark=pentagon*, mark size=1.8pt, thick} % Pentagon
}

\definecolor{c0}{RGB}{0,0,0}       % Black
\definecolor{c1}{RGB}{230,159,0}   % Orange
\definecolor{c2}{RGB}{86,180,233}  % Sky Blue
\definecolor{c3}{RGB}{240,228,66}  % Yellow
\definecolor{c4}{RGB}{0,114,178}   % Blue
\definecolor{c5}{RGB}{213,94,0}    % Vermillion
\definecolor{c6}{RGB}{204,121,167} % Reddish Purple
\definecolor{c7}{RGB}{73,0,146}    % Deep Violet
\definecolor{c8}{RGB}{146,73,0}    % Brown
\definecolor{c9}{RGB}{153,153,153} % Grey

\pgfplotsset{
    colormap={classmap}{
        color(0cm)=(c0); color(1cm)=(c1); color(2cm)=(c2); 
        color(3cm)=(c3); color(4cm)=(c4); color(5cm)=(c5); 
        color(6cm)=(c6); color(7cm)=(c7); color(8cm)=(c8); 
        color(9cm)=(c9);
    }
}

\begin{document}

% Good Representations Do Not Imply Good Retrieval
\title{Geometric Analysis of Self-Supervised Vision Representations for Semantic Image Retrieval}

\author {
\IEEEauthorblockN{Esteban Rodríguez-Betancourt}
\IEEEauthorblockA{Posgrado en Computación e Informática\\
Universidad de Costa Rica\\
esteban.rodriguezbetancourt@ucr.ac.cr}

\and

\IEEEauthorblockN{Edgar Casasola-Murillo}
\IEEEauthorblockA{
Escuela de Ciencias de la Computación\\
Universidad de Costa Rica\\
edgar.casasola@ucr.ac.cr}
}

\maketitle

\begin{abstract}
Content-based image retrieval (CBIR) systems enable users to search images based on visual content instead of relying on metadata. The text domain has benefited from vector search of representations created with unsupervised methods such as BERT. However, modern self-supervised learning methods for vision are mostly not reported in CBIR-related literature, instead relying on supervised models or multi-modal methods that align text and vision.

We evaluate how the representations learned by modern self-supervised learning methods for vision perform under typical retrieval stacks that leverage vector databases and nearest neighbor search. Our evaluation reveals that the latent space geometry impacts approximate nearest neighbor (ANN) indexing. Specifically, highly anisotropic representations with high skewness produced by several modern SSL methods degrade the performance of partition-based and hashing-based search, even if their own linear probe or K-NN accuracy is not affected. In contrast, representations with higher isotropy and local purity better satisfy the distance-based assumptions of ANN indexes, leading to improved semantic retrieval performance.
\end{abstract}


\section{Introduction}
Modern self-supervised methods such as SimCLR \cite{chen2020simclr} have been widely successful in representation learning. Despite this success, modern pure vision self-supervised methods are only sparsely mentioned in content-based image retrieval (CBIR) related literature. In this domain, methods that are supervised on labels or textual cues are usually preferred.

Vector retrieval typically relies on using a vector as a query and finding the $K$ nearest neighbors to be used as the result set. Modern retrieval systems also rely on advanced re-ranking of results; however, for this paper, we focus on nearest neighbor search. Vector search assumes that similar entities will have similar vector representations. Unfortunately, being similar in latent space does not necessarily mean being aligned to concepts a user would care about.

SSL embeddings are generally evaluated using linear probes or KNN. However, strong performance under these evaluation protocols does not necessarily translate to good retrieval performance under approximate nearest neighbor indexing. For practitioners engineering retrieval systems, this mismatch between typical evaluations and behavior on real retrieval systems can cause increasing costs, more complex postprocessing pipelines, or complete failure of the system.

Content-based image retrieval can be divided into three different subtasks: instance-level, category-level, and fine-grained image retrieval. Instance-level retrieval consists of finding the same physical object or place, just from a different viewpoint, lighting, or partial occlusion. Category-level, also known as semantic image retrieval, focuses on finding images that belong to the same semantic class or category, independent of being the same physical instance. Lastly, fine-grained image retrieval focuses on finding instances of the same class, but in this case the classes can be much more similar between them, for instance, models of cars, species of birds, or types of food. In this paper we will focus exclusively on category-level (ImageNet-1k) and fine-grained image retrieval (Food-101).

% 5. Our contribution!
The purpose of this work is to investigate how self-supervised methods perform on semantic-level retrieval tasks and explore the underlying reasons for their performance. We evaluate eight pure-vision self-supervised representation learning methods: SimCLR \cite{chen2020simclr}, DINO \cite{caron2021dino}, Barlow Twins \cite{Zbontar2021}, BYOL \cite{grill2020bootstrap}, VICReg \cite{bardes2022vicreg}, ReSA \cite{weng2025resa}, Hypersolid \cite{rodriguezbetancourt2026hypersolidemergentvisionrepresentations}, and a supervised baseline. We trained those methods on the Food-101 \cite{bossard14} and ImageNet-1k \cite{deng2009imagenet} datasets and evaluated the produced embeddings across three common approximate nearest neighbor (ANN) indexes: IVF, HNSW, and LSH. Additionally, we contribute a detailed analysis of the emergent latent space geometries. We show that the topological structure of these representations affects how suitable they are for usage in vector search pipelines by measuring embedding anisotropy, LSH bucket entropy, unsupervised clustering density, and local class purity decay.

Our contributions are fourfold:
\begin{enumerate}
    \item We provide a systematic evaluation of several modern self-supervised vision representations for semantic image retrieval across multiple ANN indexing methods (IVF, HNSW, and LSH).
    \item We show that strong performance under standard SSL evaluation protocols (e.g., linear probe and K-NN) does not necessarily translate to good retrieval performance.
    \item We provide a geometric analysis associating anisotropy, skewness, cluster structure, and neighborhood purity to the observed retrieval behavior.
    \item We provide takeaways for practitioners and researchers to improve retrieval systems and specialize SSL methods for semantic retrieval.
\end{enumerate}

It is not a goal of this paper to prescribe a method that is superior to others; instead, we want to study which properties of the geometry of the latent space induced by SSL methods help out in semantic retrieval. Additionally, we purposely used minimal setups (for instance, IVF with a single probe) to focus on the impact of the latent space geometry on retrieval. It is not a goal of this paper to provide a full semantic retrieval stack; that would require properly tuned indexes, re-rankers, and additional discriminative components.

The remainder of this paper is structured as follows. \Cref{sec:RelatedWork} reviews related work. \Cref{sec:ExperimentalSetup} describes the experimental setup. \Cref{sec:Results} presents the retrieval results. \Cref{sec:Analysis} analyzes the geometry of the learned representations. \Cref{sec:Discussion} discusses the implications of our findings, and \Cref{sec:Conclusions} concludes the paper.

\section{Related Work}\label{sec:RelatedWork}

% 1. The Broad Context & The Gap in Literature
Historically, CBIR-related methods have been far better for instance-level retrieval than for category-level retrieval \cite{eakins1999content,qazanfari2023advancementscontentbasedimageretrieval}. The gap between both capacities has been denominated as the semantic gap. Parallel to this, the vision domain has seen a surge in self-supervised learning (SSL) methods, such as \cite{caron2021dino}, which have demonstrated a better ability to learn representations with high semantic alignment. Despite these advancements, modern pure-vision self-supervised learning methods remain surprisingly scarce in content-based image retrieval literature. On the other hand, SSL research typically evaluates itself using linear probe accuracy, KNN accuracy, or fine-tuning accuracy, which do not necessarily translate to good retrieval performance. Some papers, such as DINO \cite{caron2021dino}, report results on copy detection tasks (a kind of instance-level image retrieval), but in general most SSL research does not report performance on retrieval tasks.

% 2. The Dominance of Multimodal and the "Why"
The scarcity of SSL methods in recent surveys on content-based image retrieval suggests that most retrieval research may not be leveraging these modern pure-vision methods \cite{qazanfari2023advancementscontentbasedimageretrieval,Zhang2024cbirSurvey,zheng2025retrievalaugmentedgenerationunderstanding}. \cite{dubey2022cbirdeeplearning} offers a particularly rich survey of deep-learning contributions to image retrieval; however, most of the self-supervised methods mentioned precede SimCLR and the rest of modern SSL methods. Overall, the field relies heavily on domain-specific supervised methods or multimodal (text and vision) approaches. Within the multimodal category, the CLIP family is widely mentioned in recent CBIR literature. While these models tend to perform exceptionally well, they require training on aligned text and images, making them unsuitable for domains where such paired data is unavailable. Consequently, it is unclear whether the lack of pure-vision SSL in CBIR is due to an inherent flaw in these methods for retrieval tasks or simply because most modern SSL architectures emerged after 2019.

% 3. What HAS been done (Instance-level and Domain-specific)
While self-supervised methods have been applied to image retrieval, the focus has largely remained on instance-level tasks like landmark or place identification \cite{qazanfari2023advancementscontentbasedimageretrieval,Zhang2024cbirSurvey,zheng2025retrievalaugmentedgenerationunderstanding}. For example, \cite{krishna2021earlycontrastive_eval} compares SimCLR, MoCo, AMDIM, and a supervised baseline. Furthermore, self-supervised representation learning methods have been proposed that are optimized explicitly for image retrieval. For instance, InsCLR \cite{deng2022insclr} leverages instance-level contrast by mining meaningful pseudo-positive samples. Similarly, VG-SSL \cite{Xiao2025vgssl} presents a new self-supervised method focused on geo-localization, comparing it against several modern self-supervised methods. However, in both cases, the evaluation is done in landmark related retrieval, and the methods are not evaluated in semantic-level retrieval. Beyond instance-level tasks, self-supervised methods have also been explored in specific domains. In content-based image retrieval in the radiology domain \cite{DENNER2025110640}, DINO and MAE were evaluated against other supervised, weakly supervised, and specialist methods. In this domain, self-supervised methods performed similarly or better than supervised and weakly supervised methods, though all evaluated methods performed worse than specialized models.

The same surveys mention that the semantic gap remains a challenge for CBIR systems \cite{qazanfari2023advancementscontentbasedimageretrieval}. Currently this gap is approached by using relevance feedback and re-ranking post-processing \cite{qazanfari2023advancementscontentbasedimageretrieval}.

% 4. The Geometric Perspective
From a theoretical self-supervised learning perspective, alignment and uniformity on the hypersphere have been identified as necessary properties for good quality embeddings \cite{wangisola2020}. However, geometries that perform well for classification using linear probes or K-NN do not necessarily generalize well for retrieval. Recent work by \cite{chung2026globalgeometryvisionrepresentations} shows that different methods produce geometries that bind to different features. While all methods can correctly identify parts of the image (presence of certain shapes), not all methods preserve the right layout (the right shape located in the same place). This is related to the findings of \cite{luthra2026directionalneuralcollapseexplains}, which identified that many SSL methods produce highly anisotropic embeddings and that this anisotropy is necessary for feature re-usage across many classes. However, as we show here, that anisotropy and feature re-usage make semantic retrieval much harder.

\section{Experimental Setup}\label{sec:ExperimentalSetup}
We used a mix of pretrained networks for ImageNet-1k and our own trained networks for Food-101. For ImageNet-1k, we used ResNet-50 \cite{he2016resnet} pretrained for 100 epochs by the Lightly Framework team, in addition to the 100 epoch batch size of 256 of ReSA and our own trainings of a supervised baseline and Hypersolid (100 epochs, batch size 512, AdamW with LR=$10^{-3}$ and mixed bfloat16 precision). For Food-101, we trained a ResNet-18 \cite{he2016resnet} on Food-101 \cite{bossard14} for 200 epochs, using a batch size of 128, AdamW (LR=$10^{-3}$) and mixed bfloat16 precision. As a reference, \Cref{tab:accuracy-table} summarize the performance of each model when evaluated with a linear probe or KNN ($K=200$ for ImageNet-1k and $K=224$ for Food-101).

\begin{table}[thbp]
\caption{Reported Linear Probe and KNN Accuracy}\label{tab:accuracy-table}
\begin{center}
    \begin{small}
      \begin{sc}
\begin{tabular}{lllccc}
\hline
    & &
    \multicolumn{2}{c}{\bfseries Linear} & 
    \multicolumn{2}{c}{\bfseries K-NN} \\
    \multicolumn{1}{c}{\bfseries } & 
    \multicolumn{1}{c}{\bfseries Method} & 
    \multicolumn{1}{c}{\bfseries Top 1} & \multicolumn{1}{c}{\bfseries Top 5} &
    \multicolumn{1}{c}{\bfseries Top 1} & \multicolumn{1}{c}{\bfseries Top 5} \\
    \hline
\multirow{9}{*}{\rotatebox[origin=c]{90}{\bfseries \;ImageNet-1k\;}}
    & \textcolor{gray}{Supervised}        & \textcolor{gray}{70.02} & \textcolor{gray}{89.34} & \textcolor{gray}{61.72} & \textcolor{gray}{85.29} \\
    & SimCLR      & 62.03 & 84.75 & 45.60 & 74.18 \\
    & BYOL        & 61.84 & 84.64 & 45.88 & 74.61 \\
    & SwAV        & \textbf{66.15} & \textbf{87.84} & 50.19 & 78.93 \\
    & Barlow Twins          & 59.69 & 82.68 & 46.24 & 74.44 \\
    & VICReg      & 62.66 & 84.93 & 47.14 & 75.71 \\
    & DINO        & 64.43 & 85.85 & 51.28 & 79.93 \\
    & ReSA        & 60.85 & 83.91 & \textbf{60.79} & \textbf{85.58} \\
    & Hypersolid  & 55.59 & 81.29 & 51.14 & 78.46 \\
    \hline
    \multirow{7}{*}{\rotatebox[origin=c]{90}{\bfseries Food-101}}
    & Supervised  & \textcolor{gray}{71.33} & \textcolor{gray}{89.59} & \textcolor{gray}{70.01} & \textcolor{gray}{90.35} \\
    & SimCLR      & 61.15 & 85.07 & 49.99 & 77.45 \\
    & BYOL        & 58.21 & 82.75 & 45.83 & 73.85 \\
    & Barlow Twins          & 64.11 & 86.66 & 54.90 & 80.15 \\
    & VICReg      & 65.69 & 87.34 & 56.61 & 81.19 \\
    & DINO        & 64.48 & 87.15 & 55.43 & 81.60 \\
%    & LeJEPA     & 61.86 & 85.65 & 53.76 & 79.40 \\
    & Hypersolid & \textbf{71.32} & \textbf{90.57} & \textbf{64.27} & \textbf{86.57} \\
    \hline
\end{tabular}
\end{sc}
\end{small}
\end{center}
\end{table}

After training the models, we extracted feature embeddings for each image in the validation set. We compared three different indexing algorithms to evaluate retrieval quality, using the FAISS library \cite{douze2024faiss}. The first evaluated method was IVF, implemented using the IndexFlatL2 quantizer and a list size of 100. The second one was HNSW, implemented using IndexHNSWFlat over L2 normalized embeddings, using an inner product metric, a maximum number of neighbors per node of 16, a dynamic candidate list of 40, and a search candidate list size of 16. At last, we used IndexLSH with 128 bits.

For measuring retrieval quality, we used the standard metrics Precision@10 (P@10), Recall@10 (R@10), Mean Average Precision at 10 (MAP@10), and Mean Reciprocal Rank (MRR). In all cases, higher values are considered better.

Precision@10 measures the fraction of the top-10 retrieved images that belong to the same class as the query image. Recall@10 measures the proportion of relevant images for the query that appear within the top-10 retrieved results. MAP@10 summarizes the ranking quality by computing the average precision over the top-10 retrieved items for each query and then averaging across all queries. MRR evaluates how early the first relevant item appears in the ranked list by averaging the reciprocal of the rank of the first correct retrieval. In all experiments, the images belonging to the same class as the query are considered relevant.

To analyze how the embedding geometry affects the retrieval results, we performed several evaluations: measure anisotropy, skewness, worst case hub, analyze LSH bucket occupancy vs entropy, use DBSCAN clustering to probe the geometry and its organization and measure local class purity.

Skewness measures whether nearest-neighbor retrieval is evenly distributed across items or dominated by a small number of repeated hubs. Those hubs may ``contaminate'' the purity of the retrieved neighborhood or affect the routing in HNSW style indexes.

Anisotropy is the property of non-uniformity in different directions. In latent spaces, anisotropy shows itself as embeddings crowding a narrow cone, rather than being distributed uniformly across the latent space. For SSL representations, it is known that uniformity (in this sense equivalent to high isotropy) is a desired property \cite{wangisola2020}; however, recent research shows that not necessarily highly anisotropic geometries lack class separations \cite{luthra2026directionalneuralcollapseexplains}. To understand the macro-level structure of the evaluated geometries, we measure the anisotropy of the latent spaces produced by each model.

Anisotropy breaks design assumptions of vector indexing methods, such as LSH, that expect isotropic latent spaces to properly balance load across each generated bucket. To quantify the anisotropy impact on LSH performance, we performed a stress test by projecting the embeddings into a restricted 16-bit LSH space, measured the Shannon entropy of the resulting distribution, and measured the percent of data assigned to the most filled bucket. In this regard, more buckets and higher entropy should improve LSH filtering speed, while skewed data distribution across buckets impacts performance, as the index loses its capacity to filter out irrelevant elements.

Good semantic retrieval also requires semantically related elements to be close to each other. To measure this, we used the DBSCAN clustering algorithm, both using cosine and Euclidean distances. DBSCAN is unique among clustering algorithms as it does not need to enforce a number of clusters; instead, it discovers related nodes by how close they are to their neighbors. Therefore, DBSCAN can signal if similar items are organized closely and also determine if there are space gaps between semantically unrelated items.

Finally, we measured local purity decay, which consists of determining the percent of the nearest K neighbors that belong to the same class as a given embedding, as a function of K. This analysis allows us to determine how homogeneous the geometry is class-wise and determine how fast the neighborhood ends up mixing up different classes. This impacts the perceived quality of a retrieval system, as the initial vector query will contain more semantically related items, reducing the amount of false positives to filter away or reduce noise perceived by the user.

\section{Results}\label{sec:Results}
The retrieval performance on ImageNet-1k and Food-101 is summarized in \Cref{tab:RetrievalImageNet}. For each dataset, we used each indexed item itself as a query.

\newcommand{\fmtvalue}[2]{$#1_{\color{gray}#2}$}
\newcommand{\bestfmtvalue}[2]{$\textbf{#1}_\textbf{\color{gray}#2}$}

\begin{table*}[tbhp]
\caption{Retrieval Metrics on ImageNet-1k across ANN index types (mean$\times 100$ and std$\times 100$)}
\begin{center}
\begin{tabular}{|ll|cccc|cccc|cccc|}
\hline
& \textbf{Method} &
\multicolumn{4}{c|}{\textbf{IVF (nprobe=1)}} &
\multicolumn{4}{c|}{\textbf{HNSW Index}} &
\multicolumn{4}{c|}{\textbf{LSH Index (128-bit Hashing)}} \\
 \cline{3-14}
&& \textbf{P@10} & \textbf{R@10} & \textbf{mAP@10} & \textbf{MRR}
& \textbf{P@10} & \textbf{R@10} & \textbf{mAP@10} & \textbf{MRR}
& \textbf{P@10} & \textbf{R@10} & \textbf{mAP@10} & \textbf{MRR} \\
\hline

\multirow{9}{*}{\rotatebox[origin=c]{90}{\bfseries \;ImageNet-1k\;}}
\input{data/indexes_metrics_im1k}

\\\hline
\multirow{7}{*}{\rotatebox[origin=c]{90}{\bfseries \;Food-101\;}}
\input{data/indexes_metrics_food101}
\\\hline
\end{tabular}%
\label{tab:RetrievalImageNet}
\end{center}
\end{table*}

To better understand the factors behind these results, we analyze the geometric structure of the learned embedding spaces in the following section.

\section{Analysis of Latent Space Geometry}\label{sec:Analysis}

Different training methods produce vastly different latent space geometries. Retrieval precision and recall depend on how this geometry is organized. To improve retrieval results, it is important that images of the same class are clustered together around a small area and that those clusters are separated from other clusters by wide gaps. Finally, those clusters must be aligned with the classes we expect; otherwise, we would retrieve unexpected images.

\subsection{Geometry}

\newcommand{\topologyVizWidth}{0.32\textwidth}
\begin{figure*}[!t]
    \centering
    \subfloat[Supervised]{\parbox[t]{\topologyVizWidth}{\centering
        \includegraphics[width=\topologyVizWidth]{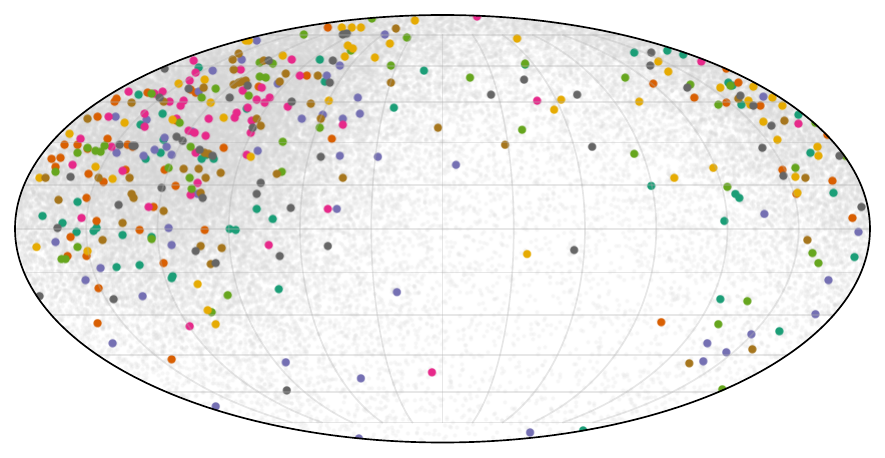}}}
    \subfloat[SimCLR]{\parbox[t]{\topologyVizWidth}{\centering
        \includegraphics[width=\topologyVizWidth]{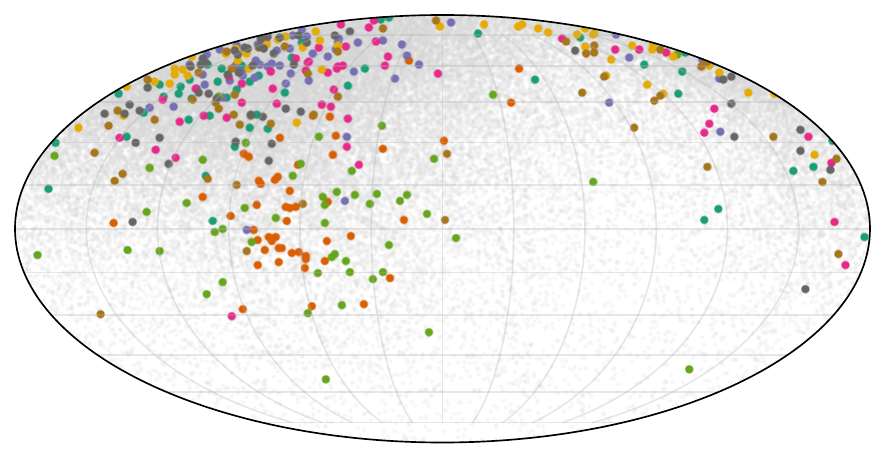}}}
    \subfloat[BYOL]{\parbox[t]{\topologyVizWidth}{\centering
        \includegraphics[width=\topologyVizWidth]{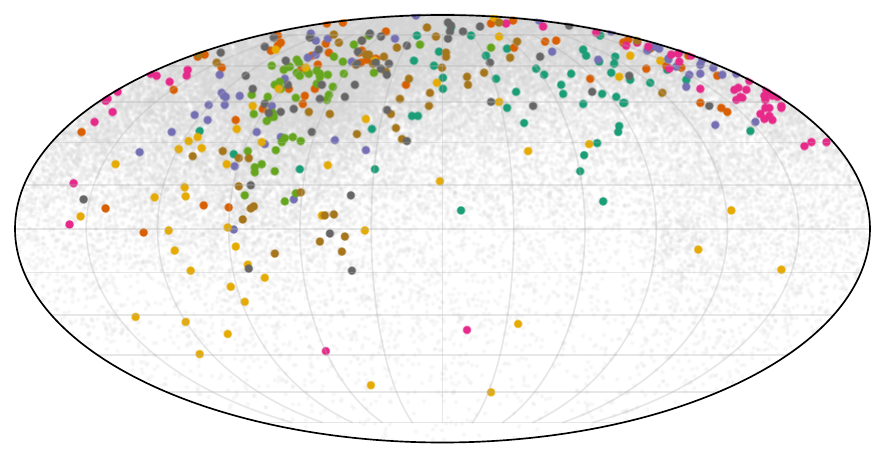}}}\\[0.5em]
    \subfloat[SwAV]{\parbox[t]{\topologyVizWidth}{\centering
        \includegraphics[width=\topologyVizWidth]{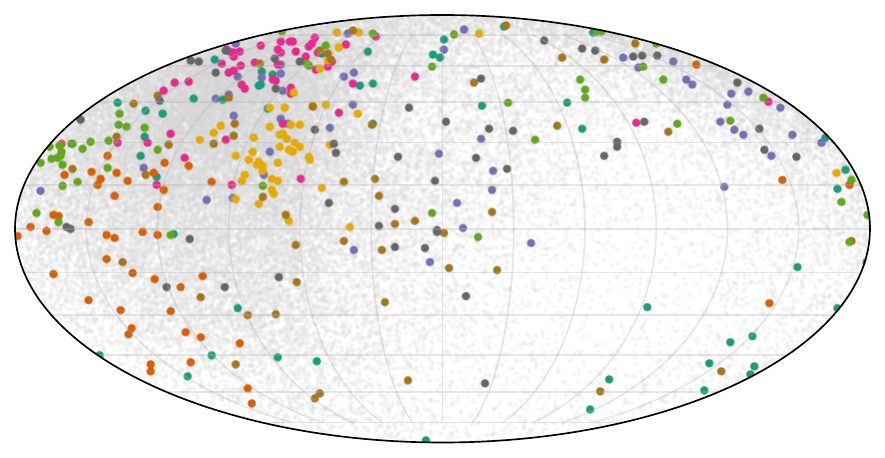}}}        
    \subfloat[Barlow Twins]{\parbox[t]{\topologyVizWidth}{\centering
        \includegraphics[width=\topologyVizWidth]{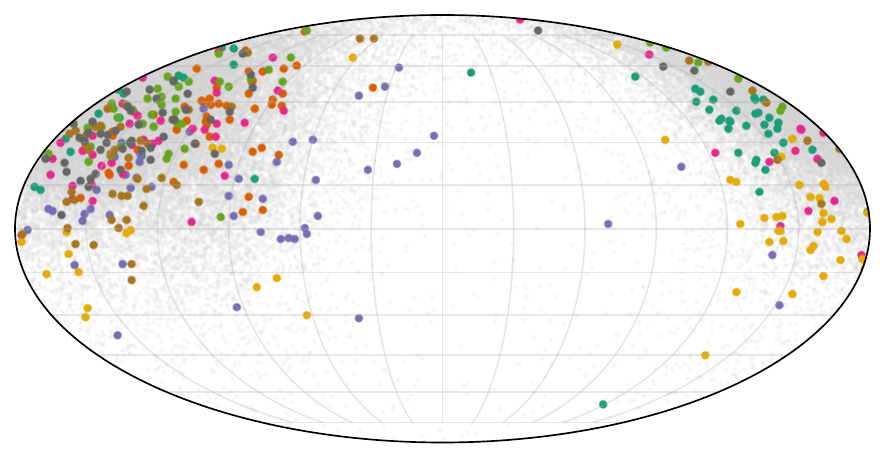}}}
    \subfloat[VICReg]{\parbox[t]{\topologyVizWidth}{\centering
        \includegraphics[width=\topologyVizWidth]{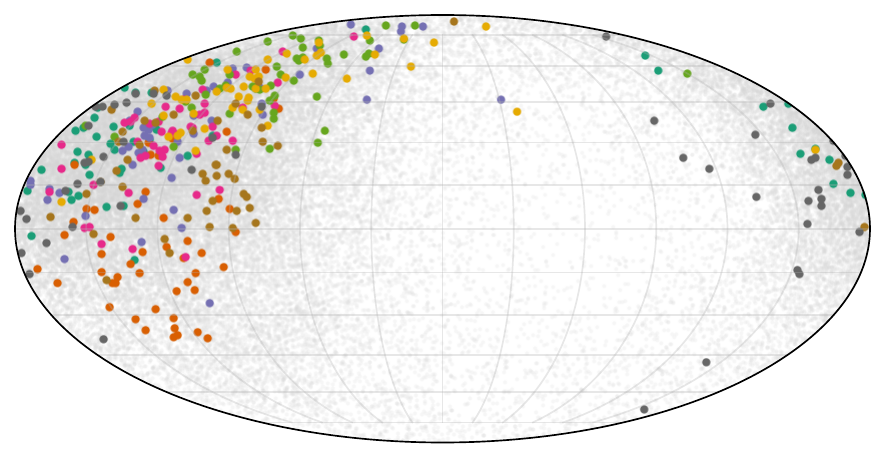}}}\\[0.5em]
    \subfloat[DINO]{\parbox[t]{\topologyVizWidth}{\centering
        \includegraphics[width=\topologyVizWidth]{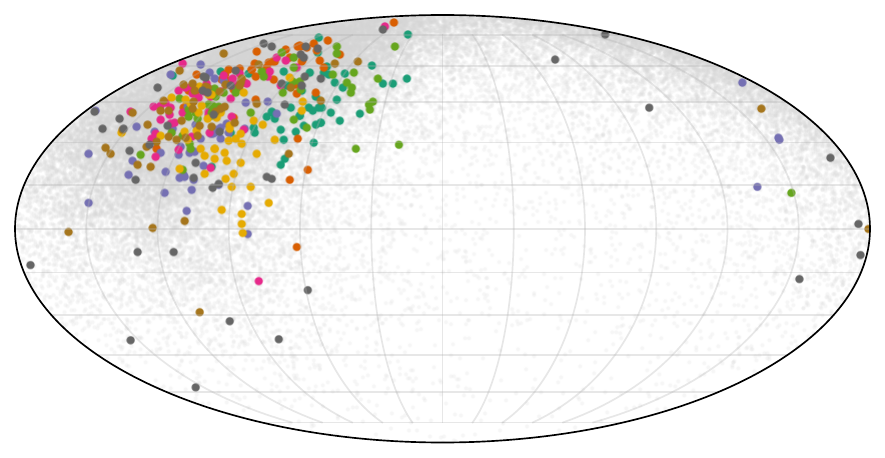}}}
%    \subfloat[LeJEPA]{\parbox[t]{\topologyVizWidth}{\centering
%        \includegraphics[width=\topologyVizWidth]{data/topology_mollweide/mollweide_lejepa_resnet50_imagenet.pdf}}}
    \subfloat[ReSA]{\parbox[t]{\topologyVizWidth}{\centering
        \includegraphics[width=\topologyVizWidth]{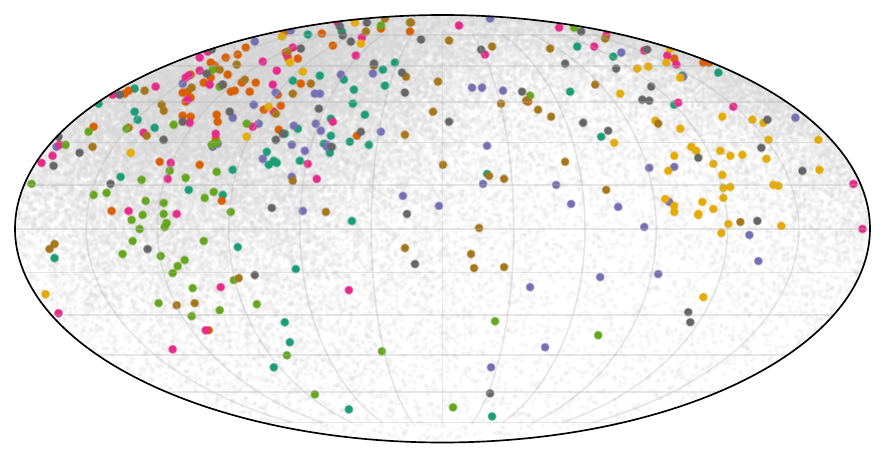}}}
    \subfloat[Hypersolid]{\parbox[t]{\topologyVizWidth}{\centering
        \includegraphics[width=\topologyVizWidth]{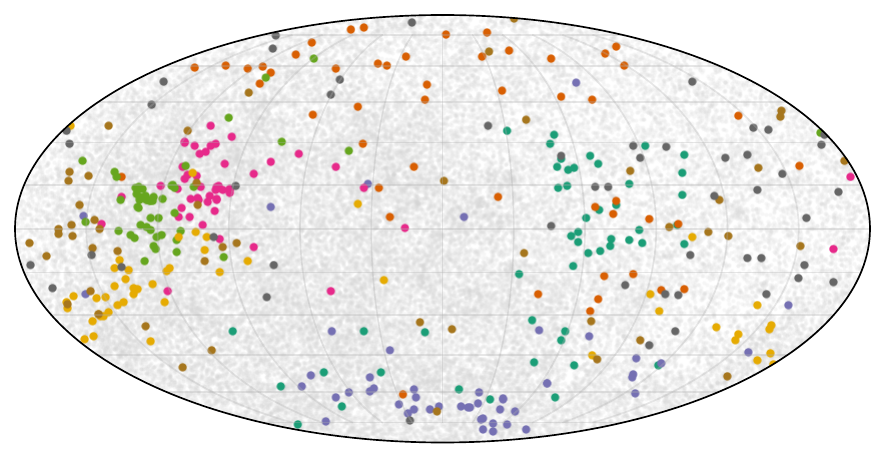}}}

    \caption{Latent space geometry visualization across different learned representation methods on ImageNet-1k}
    \label{fig:manifold_topologies}
\end{figure*}

A desirable property in good representations is alignment and uniformity \cite{wangisola2020}. For retrieval, those properties help the index in different ways. Alignment provides spatial locality, assigning close regions of the latent space to, ideally, semantically related elements. Then, an index that retrieves the closest neighbors would necessarily retrieve semantically related elements. On the other hand, uniformity contributes by better using the latent space. This manifests as bigger margins between elements and classes, making it easier for the index to balance the load in whatever structure it uses to split the latent space.

Mathematically, the previous intuition can be approached by measuring the latent space by \emph{anisotropy}, \emph{skewness,} and \emph{hubness}. These measurements are summarized in \Cref{tab:anisotropy}.

\begin{table}[htbp]
\caption{Embedding Anisotropy and Skewness on ImageNet-1k}
\begin{center}
\begin{tabular}{|l|c|c|c|}
\hline
\textbf{Method} & \textbf{Anisotropy} & \textbf{Skewness} & \textbf{Worst-Case Hub} \\
\hline
Supervised   & 0.29 & 1.64 & 148\\
SimCLR       & 0.26 & 1.50 & 88\\
BYOL         & 0.35 & 1.83 & 118 \\
SwAV         & 0.24 & 1.42 & 86 \\
Barlow Twins & 0.39 & 1.41 & 93 \\
VICReg       & 0.25 & 0.98 & \textbf{61} \\
DINO         & 0.53 & 1.20 & 77 \\
%LeJEPA       & 0.47 & 1.99 & 106 \\
ReSA         & 0.21 & 1.36 & 92\\
Hypersolid   & \textbf{0.11} & \textbf{0.88} & 66\\
\hline
\end{tabular}
\label{tab:anisotropy}
\end{center}
\end{table}

\emph{Anisotropy} measures the concentration of variance along the dominant directions in the embedding space. It is defined as the fraction of total variance explained by the leading principal component:

\[
\text{Anisotropy} = \frac{\lambda_{\max}}{\text{Tr}(\Sigma)}
\]

where $\Sigma$ is the covariance matrix of the embeddings, and $\lambda_{\max}$ is its largest eigenvalue. Low anisotropy means variance is spread across many directions and that the representation is \emph{isotropic}. On the other hand, higher anisotropy means that variance is mostly concentrated in fewer directions, leading to collapse or cone-like structures.

The next identified metric is \emph{skewness}. Let $N_k(j)$ be the number of times that item $j$ appears in the top-k nearest neighbor list over all queries, for $j=1,\ldots,n$. Let $\mu$ and $\sigma$ denote the sample mean and standard deviation of ${N_k(j)}^n_{j=1}$. The \emph{skewness} of the distribution is defined as the standardized third central moment:

\[
\text{Skewness} = \frac{n}{(n-1)(n-2)}\sum_{j=1}^n\left(\frac{N_k(j)-\mu}{\sigma}\right)^3
\]

and \emph{worst-case hubness} is defined as: \[\max_j N_k(j)\]

Lower skewness means that there are fewer or less extreme hubs, leading to more evenly distributed neighbor occurrences across items.

Our sampling size is small; however, we can point out that there is an observed inverse Spearman correlation between skewness and better retrieval metrics when using IVF and LSH indexes, as shown in \Cref{tab:full_correlation}.

\begin{table}[htbp]
\caption{Spearman correlation ($\rho$) and p-values between embedding properties and retrieval metrics}
\centering
\begin{tabular}{|c|l|cc|cc|cc|}
\hline
\textbf{Idx} & \textbf{Metric} 
& \multicolumn{2}{c|}{\textbf{Anisotropy}} 
& \multicolumn{2}{c|}{\textbf{Skewness}} 
& \multicolumn{2}{c|}{\textbf{Max Hub}} \\
\cline{3-8}
& & $\rho$ & p & $\rho$ & p & $\rho$ & p \\
\hline

\multirow{4}{*}{\rotatebox{90}{IVF}}
& P@10   & -0.47 & 0.203 & \textbf{-0.67} & \textbf{0.049} & -0.40 & 0.284 \\
& R@10   & -0.47 & 0.206 & \textbf{-0.67} & \textbf{0.047} & -0.33 & 0.389 \\
& mAP@10 & -0.47 & 0.203 & \textbf{-0.67} & \textbf{0.049} & -0.40 & 0.284 \\
& MRR    & -0.44 & 0.232 & -0.54 & 0.137 & -0.24 & 0.529 \\
\hline

\multirow{4}{*}{\rotatebox{90}{HNSW}}
& P@10   & -0.53 & 0.145 & -0.40 & 0.284 & -0.13 & 0.731 \\
& R@10   & -0.47 & 0.197 & -0.56 & 0.117 & -0.24 & 0.539 \\
& mAP@10 & -0.54 & 0.135 & -0.43 & 0.250 & -0.24 & 0.539 \\
& MRR    & -0.54 & 0.135 & -0.43 & 0.250 & -0.24 & 0.539 \\
\hline

\multirow{4}{*}{\rotatebox{90}{LSH}}
& P@10   & -0.41 & 0.273 & \textbf{-0.80} & \textbf{0.009} & \textbf{-0.74} & \textbf{0.024} \\
& R@10   & -0.41 & 0.273 & \textbf{-0.74} & \textbf{0.024} & \textbf{-0.74} & \textbf{0.024} \\
& mAP@10 & -0.41 & 0.273 & \textbf{-0.74} & \textbf{0.024} & \textbf{-0.74} & \textbf{0.024} \\
& MRR    & -0.41 & 0.273 & \textbf{-0.80} & \textbf{0.009} & \textbf{-0.74} & \textbf{0.024} \\
\hline

\end{tabular}
\label{tab:full_correlation}
\end{table}

To give an intuition into why these properties affect retrieval, we build a visualization of the latent space geometry in \Cref{fig:manifold_topologies}. For this visualization, we leveraged the Johnson-Lindenstrauss lemma by projecting the embeddings into a 3D sphere using a random matrix. Then we projected that sphere into 2D using a Mollweide projection. Each point was gray colored, but we colored 8 random classes (the same classes for each model, colored with the same color).

Higher anisotropy implies that embeddings concentrate along a few dominant directions, reducing effective usage of the latent space and leading to more ``clumped'' representations with potentially higher inter-class overlap. Lower skewness of the $N_k$ distribution indicates a more even distribution of neighbor occurrences.

This spatial concentration actively degrades the efficacy of approximate nearest neighbor indexes. Graph-based methods such as HNSW \cite{malkov2020hnsw} are built on the assumption that distances are highly discriminative, allowing graph jumps to discard most candidates quickly. Partition-based methods such as the Inverted File index (IVF) require partitioning the space into balanced Voronoi cells, an optimization that becomes highly inefficient if the majority of vectors are heavily concentrated in a narrow geometric region \cite{jegou2011pqnns}. Finally, Locality Sensitive Hashing (LSH) relies on the assumption that elements are sufficiently spread across the latent space to be separated into distinct, manageable buckets. While LSH requires similar elements to cluster together locally to maximize collision probability \cite{indyk1998lsh}, high global anisotropy actively degrades the hash functions by drastically reducing the number of effective buckets while inflating the size of the few utilized ones \cite{dong2008lshmodeling}.

\subsection{LSH Bucket Analysis}
LSH indexes are particularly attractive at scale because they allow $O(1)$ routing between an embedding and the ``shard'' that should contain it, if any. However, for an efficient deployment, it is required that buckets are evenly balanced.

To exemplify how different models can distribute the embeddings unevenly in the latent space, we built an LSH index using 16 bits, leading to a maximum theoretical entropy of 15.61 bits (there is a maximum of $2^{16}$ buckets, but the validation set has just $50000$ images: $15.61\approx\log_2 50000$). Some models like BYOL or DINO ended up with fewer unique buckets and lower entropy, leading to some buckets having many more elements, as shown in \Cref{tab:lsh_entropy}.

\begin{table}[htbp]
\caption{LSH Bucket Occupancy and Entropy (n=16 bits) on ImageNet-1k}
\begin{center}
\begin{tabular}{|l|c|c|c|}
\hline
\textbf{Method} & \textbf{Unique Buckets} & \textbf{Entropy} & \textbf{Max Bucket \%} \\
\hline
Supervised   & 10614 & 11.84 & 0.55\%  \\
SimCLR       & 9659  & 11.65  & 1.06\%  \\
BYOL         & 4320  & 9.64  & 2.53\%  \\
SwAV         & 11558 & 12.11 & 0.77\% \\
Barlow Twins & 6816   & 11.05  & 0.68\% \\
VICReg       & 9133  & 11.53  & 0.98\%  \\
DINO         & 2314 & 8.27 & 3.14\%  \\
%LeJEPA       & 5730  & 10.28 & 2.13\%  \\
ReSA         & \textbf{19749} & \textbf{13.62} & \textbf{0.12}\% \\
Hypersolid   & 14056 & 12.88 & 0.22\%  \\
\hline
\end{tabular}
\label{tab:lsh_entropy}
\end{center}
\end{table}

But what determines whether an embedding distribution is suitable for LSH indexes? As shown in \Cref{tab:lsh_occupancy_correlations}, in this case we need to look at \emph{anisotropy}. Higher anisotropy leads to less unique buckets, lower entropy, and bigger buckets. From a performance point of view, unevenly distributed, bigger buckets are not desirable because LSH performs a linear scan within the bucket to find the closest matches. Huge buckets that may contain many classes are not discriminative and require much more processing.

\begin{table}[htbp]
\caption{Spearman correlation ($\rho$) and p-values between LSH Bucket Properties and Embedding Properties}
\centering
\begin{tabular}{|l|cc|cc|cc|}
\hline
\textbf{Metric} 
& \multicolumn{2}{c|}{\textbf{Anisotropy}} 
& \multicolumn{2}{c|}{\textbf{Skewness}} 
& \multicolumn{2}{c|}{\textbf{Max Hub}} \\
\cline{2-7}
& $\rho$ & p & $\rho$ & p & $\rho$ & p \\
\hline
Unique Buckets & \textbf{-0.90} & \textbf{0.001} & -0.20 & 0.606 & -0.07 & 0.865 \\
Entropy & \textbf{-0.90} & \textbf{0.001} & -0.20 & 0.606 & -0.07 & 0.865 \\
Max Bucket \% & \textbf{0.67} & \textbf{0.050} & 0.25 & 0.516 & -0.13 & 0.732 \\\hline
\end{tabular}
\label{tab:lsh_occupancy_correlations}
\end{table}

\begin{figure}[htbp]
\centering
\begin{tikzpicture}
\begin{axis}[
    ymin=0,
    width=\columnwidth,
    xlabel={nbits (log scale)},
    ylabel={Precision@10},
    xmode=log, 
    log basis x=2, 
    log ticks with fixed point,
    xtick={16, 64, 256, 1024, 4096},
    xticklabels={16, 64, 256, 1024, 4096},
    grid=major,
    legend columns=3,
    legend style={
        at={(0.5,-0.15)},
        anchor=north,
        draw=none,
        font=\footnotesize
    },
    cycle list name=mycolorlist
]

\addplot table [x=nbits, y=Supervised_P] {data/imagenet_lsh_quality_per_nbits.txt};
\addlegendentry{Supervised}

\addplot table [x=nbits, y=SimCLR_P] {data/imagenet_lsh_quality_per_nbits.txt};
\addlegendentry{SimCLR}

\addplot table [x=nbits, y=BYOL_P] {data/imagenet_lsh_quality_per_nbits.txt};
\addlegendentry{BYOL}

\addplot table [x=nbits, y=SwAV_P] {data/imagenet_lsh_quality_per_nbits.txt};
\addlegendentry{SwAV}

\addplot table [x=nbits, y=BarlowTwins_P] {data/imagenet_lsh_quality_per_nbits.txt};
\addlegendentry{Barlow Twins}

\addplot table [x=nbits, y=VICReg_P] {data/imagenet_lsh_quality_per_nbits.txt};
\addlegendentry{VICReg}

\addplot table [x=nbits, y=DINO_P] {data/imagenet_lsh_quality_per_nbits.txt};
\addlegendentry{DINO}

%\addplot table [x=nbits, y=LeJEPA_P] {data/imagenet_lsh_quality_per_nbits.txt};
%\addlegendentry{LeJEPA}

\addplot table [x=nbits, y=ReSA_P] {data/imagenet_lsh_quality_per_nbits.txt};
\addlegendentry{ReSA}

\addplot table [x=nbits, y=HypersolidMaxPoolNorm_P] {data/imagenet_lsh_quality_per_nbits.txt};
\addlegendentry{Hypersolid}

\end{axis}
\end{tikzpicture}
\caption{LSH precision@10 on ImageNet-1k as a function of hash bits.}
\label{fig:LSHprecisionPerBit}
\end{figure}
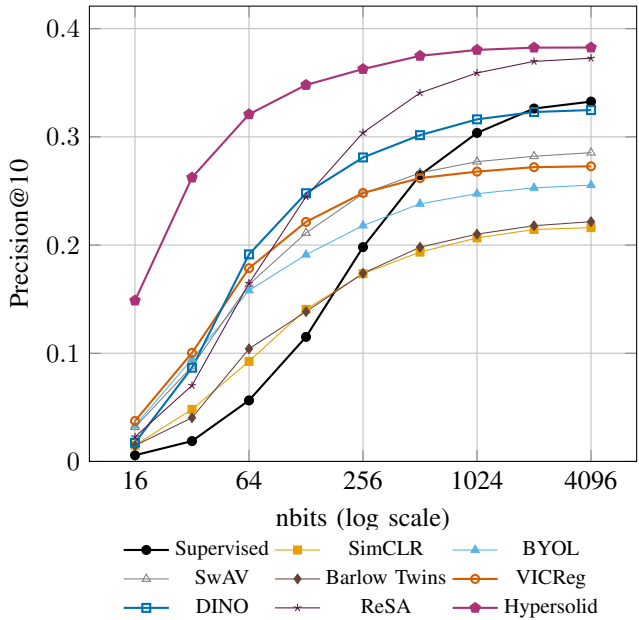

The number of unique buckets and entropy gaps can be fixed by simply increasing the number of bits used for LSH. However, even with more bits, the retrieval precision gap remains. As shown in \Cref{fig:LSHprecisionPerBit}, Hypersolid has better precision @ 10 than all other methods for each evaluated number of bits. This difference is relevant, as it not only impacts precision itself but also the storage requirements, memory consumption, and CPU usage, as smaller hashes can be queried more efficiently.

\subsection{DBSCAN Clustering}
To determine how the emergent latent space organization matches the ground truth labels, we used unsupervised clustering with DBSCAN \cite{ester1996dbscan}. This method has the advantage that it can discover by itself the number of clusters, as it relies on the distance between points to determine belonging.

\Cref{tab:DBSCANCombined} summarizes the DBSCAN results on ImageNet-1k with cosine and Euclidean distances. To make the EPS comparable across methods in the Euclidean metric case, we scaled the embeddings per method so the longest one fits within a unit hypersphere.

\begin{table}[htbp]
\caption{DBSCAN clustering performance on ImageNet-1k}
\begin{center}
\begin{tabular}{|l|ccc|ccc|}
\hline
 & \multicolumn{3}{c|}{\textbf{Cosine (eps=0.1)}} 
 & \multicolumn{3}{c|}{\textbf{Euclidean (eps=0.4)}} \\
\cline{2-7}
\textbf{Method} 
& \textbf{NMI} & \textbf{ARI} & \textbf{N}
& \textbf{NMI} & \textbf{ARI} & \textbf{N} \\
\hline
Supervised   & 0.00 & 0.0000 & 0   & 0.00 & 0.0000 & 0 \\
SimCLR       & 0.02 & 0.0000 & 45  & 0.01 & 0.0000 & 21 \\
BYOL         & 0.31 & 0.0007 & 334 & 0.23 & 0.0003 & 281 \\
SwAV         & 0.13 & 0.0001 & 199 & 0.08 & 0.0000 & 142 \\
BT           & 0.06 & 0.0000 & 109 & 0.02 & 0.0000 & 45 \\
VICReg       & 0.16 & 0.0001 & 269 & 0.10 & 0.0001 & 193 \\
DINO         & 0.21 & 0.0010 & 198 & 0.37 & 0.0012 & 387 \\
ReSA         & 0.02 & 0.0000 & 36  & 0.01 & 0.0000 & 13 \\
Hypersolid   & \textbf{0.48} & \textbf{0.0025} & \textbf{346}
              & \textbf{0.48} & \textbf{0.0017} & \textbf{422} \\
\hline
\end{tabular}
\label{tab:DBSCANCombined}
\end{center}
\end{table}

Under both metrics, ARI is generally low for all methods, indicating that cluster and class alignment is weak, if any. Hypersolid is an outlier with much higher normalized mutual information (NMI) than the other evaluated methods and many more found clusters. In the case of Hypersolid, this indicates that its embedding space contains well-separated, high-density regions weakly aligned with semantic classes. On the other hand, the supervised baseline and ReSA have a much lower NMI, which suggests vastly different geometries can still perform well on semantic retrieval tasks.

\subsection{Local Purity Decay}

\begin{figure}[tbhp]
\centering
\begin{tikzpicture}
\begin{axis}[
    width=\columnwidth,
    height=1.2\columnwidth,
    xlabel={$k$},
    ylabel={Local Purity},
    xmin=1, xmax=50,
    ymin=0.1, ymax=0.50,
    cycle list name=mycolorlist,
    grid=major,
    legend columns=3,
    legend style={
        at={(0.5,-0.1)},
        anchor=north,
        draw=none,
        font=\footnotesize
    },
    tick label style={font=\footnotesize},
    label style={font=\small},
]

\addplot table [x=k, y=Supervised] {data/local_purity.txt};
\addlegendentry{Supervised}

\addplot table [x=k, y=SimCLR] {data/local_purity.txt};
\addlegendentry{SimCLR}

\addplot table [x=k, y=BYOL] {data/local_purity.txt};
\addlegendentry{BYOL}

\addplot table [x=k, y=SwAV] {data/local_purity.txt};
\addlegendentry{SwAV}

\addplot table [x=k, y=BarlowTwins] {data/local_purity.txt};
\addlegendentry{Barlow Twins}

\addplot table [x=k, y=VICReg] {data/local_purity.txt};
\addlegendentry{VICReg}

\addplot table [x=k, y=DINO] {data/local_purity.txt};
\addlegendentry{DINO}

%\addplot table [x=k, y=LeJEPA] {data/local_purity.txt};
%\addlegendentry{LeJEPA}

\addplot table [x=k, y=ReSA] {data/local_purity.txt};
\addlegendentry{ReSA}

\addplot table [x=k, y=HypersolidMaxPoolNorm] {data/local_purity.txt};
\addlegendentry{Hypersolid}

\end{axis}
\end{tikzpicture}
\caption{Local purity as a function of neighborhood size $k$ on ImageNet-1k, using Euclidean metric.}
\label{fig:local_purity}
\end{figure}
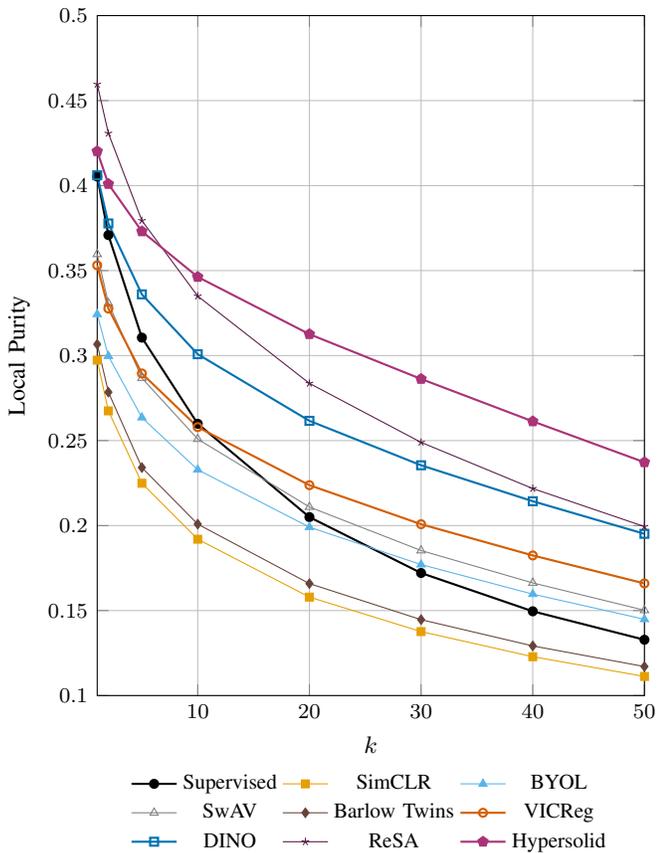

Given an embedding of a certain class, we define local purity as the percent of the nearest K neighbors that belong to the same class. \Cref{fig:local_purity} shows the evolution of purity as a function of the neighborhood size $k$. Local purity consistently decreases as $k$ increases for all methods. However, the rate of this decay varies across methods.

Hypersolid exhibits the highest local purity across all neighborhood sizes while also having a slower decay in absolute terms. This indicates that nearest neighbors in the embedding space are more likely to belong to the same semantic class, directly improving retrieval precision.

The supervised baseline also starts with a high local class purity, similar to Hypersolid at $k=1$. However, its local purity decays much faster than other methods as $k$ increases.

As it is related to local neighborhoods, local purity can be considered a rough proxy of how well an HNSW index will work in practice. HNSW indexes also benefit from embeddings following an isotropic distribution, as it allows the index to quickly discard bigger regions of the latent space.

\section{Discussion}\label{sec:Discussion}
The current investigation evaluates several self-supervised learning methods and a supervised baseline on the task of semantic retrieval. This task is particularly sensitive to the geometry of the learned latent space, as it depends on the quality of the neighborhoods.

In general, precision@10 using retrieval techniques (IVF, HNSW, or LSH) is much lower than the precision achieved using a linear probe or KNN. This makes sense, as these probes are aligned to ground truth labels, while retrieval indexes simply assume that close items must be highly similar. Vector indexing requires embeddings with a much cleaner geometry, which, as shown in this investigation, not all techniques used to learn representations achieve equally.

Most modern SSL methods are optimized to produce versatile representations usable across diverse downstream tasks. Recent literature points out that in SSL methods with highly anisotropic geometries, maintaining variance along multiple orthogonal decision axes allows models to support strong few-shot transfer and minimize task interference \cite{luthra2026directionalneuralcollapseexplains}. While preserving diverse feature axes is beneficial for a broad class of tasks, we suspect that it actively degrades the efficacy of nearest-neighbor search. ANN algorithms, such as Inverted File partitioning (IVF) and Locality Sensitive Hashing (LSH), compute distances across all dimensions uniformly. In this context, any variance that is not related to semantic relationships acts as geometric noise, causing representations to crowd into narrow regions of the hypersphere. This crowding breaks the assumptions of uniform partitioning, which is empirically evident in our LSH analysis (\Cref{tab:lsh_entropy}), where highly anisotropic models collapsed significant portions of the dataset into a single hash bucket, severely reducing retrieval precision.

Our results indicate that the multi-task geometry produced by SSL methods may not necessarily perform well in a very specific task such as semantic-level content-based image retrieval using nearest neighbors indexes. The implication for builders of image retrieval systems that want to gain semantic retrieval capabilities is that strong ``off-the-shelf'' general models may not provide the desired geometry, as they do not necessarily align with the desired semantics. For successful retrieval, it is necessary to align the representations with the desired semantics by using ground truth labels (supervised training). Therefore, it is fundamental to measure performance (in both accuracy and speed senses) and fine-tune the indexes accordingly, or even swap the model with another one (pretrained or use  specific). Another popular option is using embeddings produced by multimodal models such as CLIP. However, the same warning applies: it must be properly evaluated to confirm that the latent space geometry matches the semantics that the builder wants to enforce.

% Takeaways
For retrieval practitioners, the key takeaway is that the embeddings must be evaluated before deployment, as the linear probe accuracies or KNN hardly predict the real accuracy or recall in practice. The evaluations depend on the type of index being used. The local purity metric clearly indicates how well aligned the geometry is regarding the desired semantics. However, the isotropy must also be considered, as it is by itself an expected condition of most vector indexes. High anisotropy particularly affects methods that use $O(1)$ routing, such as LSH, causing bucket imbalances and leaving most of the bucket unused. High anisotropy also affects HNSW indexes by making it harder to exclude candidates.

In a proper retrieval system, there may also be an additional re-ranking step. This is essential to properly align the outputs with user preferences or other complex criteria. However, we can presume that a more aligned geometry can reduce the work that a re-ranking stage would have to do.

For researchers, our findings show a gap in SSL performance on semantic-level retrieval compared to typical linear probe accuracy or K-NN accuracy. This is a clear opportunity for future research, as most research is focused on producing general-purpose representations.

\section{Conclusions}\label{sec:Conclusions}

In this work, we evaluated different pure-vision self-supervised representation learning methods in the task of semantic-level content-based image retrieval and compared them with a supervised baseline. We showed that semantic retrieval does not necessarily align with typical linear probe or K-NN accuracy and that instead we should look for low anisotropy and low skewed geometries. We found that low skewness strongly correlates with higher accuracy for IVF and LSH indexes, while isotropic geometries help methods such as LSH by better distributing the data across more buckets. We also noted that while utilizing more bits can improve LSH precision, this increase does not benefit all methods equally. We also found that local purity decay varies across different methods, directly impacting the quality of the semantic performance.

In general, our findings provide an evaluation of the state of semantic-level retrieval across many modern SSL methods and provide a geometric analysis that explains the reasons behind such results. That analysis can be useful for practitioners to properly engineer their content-based image retrieval systems and can serve as guidance on which properties must be evaluated for future SSL methods. These methods may, for example, sacrifice general purpose usage and instead focus better on the needs of retrieval systems. These results can help close the semantic gap existing currently in content based image retrieval.

\bibliographystyle{ieeetr}
\bibliography{biblio}

\end{document}

%% file: data/indexes_metrics_im1k.tex
&Supervised & \fmtvalue{23}{27} & \fmtvalue{5}{5} & \fmtvalue{17}{25} & \fmtvalue{45}{44} & \fmtvalue{34}{31} & \fmtvalue{7}{6} & \fmtvalue{27}{31} & \fmtvalue{57}{43} & \fmtvalue{12}{18} & \fmtvalue{2}{4} & \fmtvalue{7}{15} & \fmtvalue{28}{39} \\
&SimCLR & \fmtvalue{16}{23} & \fmtvalue{3}{5} & \fmtvalue{11}{21} & \fmtvalue{34}{42} & \fmtvalue{22}{27} & \fmtvalue{4}{6} & \fmtvalue{16}{25} & \fmtvalue{41}{43} & \fmtvalue{14}{22} & \fmtvalue{3}{5} & \fmtvalue{10}{20} & \fmtvalue{29}{39} \\
&BYOL & \fmtvalue{21}{27} & \fmtvalue{4}{6} & \fmtvalue{15}{26} & \fmtvalue{39}{43} & \fmtvalue{25}{30} & \fmtvalue{5}{6} & \fmtvalue{19}{29} & \fmtvalue{44}{43} & \fmtvalue{19}{27} & \fmtvalue{4}{6} & \fmtvalue{14}{25} & \fmtvalue{33}{41} \\
&SwAV & \fmtvalue{22}{28} & \fmtvalue{4}{6} & \fmtvalue{16}{26} & \fmtvalue{42}{43} & \fmtvalue{29}{31} & \fmtvalue{6}{6} & \fmtvalue{22}{30} & \fmtvalue{48}{44} & \fmtvalue{21}{28} & \fmtvalue{4}{6} & \fmtvalue{15}{26} & \fmtvalue{38}{42} \\
&Barlow Twins & \fmtvalue{17}{24} & \fmtvalue{4}{5} & \fmtvalue{12}{22} & \fmtvalue{36}{42} & \fmtvalue{22}{27} & \fmtvalue{5}{6} & \fmtvalue{16}{25} & \fmtvalue{43}{43} & \fmtvalue{14}{22} & \fmtvalue{3}{4} & \fmtvalue{9}{19} & \fmtvalue{29}{39} \\
&VICReg & \fmtvalue{23}{29} & \fmtvalue{5}{6} & \fmtvalue{17}{27} & \fmtvalue{42}{43} & \fmtvalue{27}{31} & \fmtvalue{6}{6} & \fmtvalue{21}{30} & \fmtvalue{46}{43} & \fmtvalue{22}{29} & \fmtvalue{4}{6} & \fmtvalue{16}{27} & \fmtvalue{38}{42} \\
&DINO & \fmtvalue{27}{30} & \fmtvalue{5}{6} & \fmtvalue{21}{29} & \fmtvalue{47}{44} & \fmtvalue{32}{33} & \fmtvalue{7}{7} & \fmtvalue{26}{33} & \fmtvalue{52}{44} & \fmtvalue{25}{30} & \fmtvalue{5}{6} & \fmtvalue{18}{29} & \fmtvalue{41}{42} \\
&ReSA & \fmtvalue{29}{31} & \fmtvalue{6}{6} & \fmtvalue{23}{30} & \bestfmtvalue{51}{44} & \bestfmtvalue{38}{34} & \bestfmtvalue{8}{7} & \bestfmtvalue{31}{34} & \bestfmtvalue{59}{43} & \fmtvalue{24}{29} & \fmtvalue{5}{6} & \fmtvalue{18}{27} & \fmtvalue{43}{43} \\
&Hypersolid & \bestfmtvalue{32}{34} & \bestfmtvalue{7}{7} & \bestfmtvalue{26}{34} & \fmtvalue{49}{44} & \fmtvalue{37}{36} & \bestfmtvalue{8}{7} & \bestfmtvalue{31}{36} & \fmtvalue{53}{44} & \bestfmtvalue{35}{36} & \bestfmtvalue{7}{7} & \bestfmtvalue{28}{35} & \bestfmtvalue{49}{44}

%% file: data/indexes_metrics_food101.tex
&Supervised & \fmtvalue{47}{40} & \bestfmtvalue{2}{2} & \fmtvalue{41}{41} & \fmtvalue{60}{44} & \fmtvalue{52}{37} & \bestfmtvalue{2}{1} & \fmtvalue{44}{39} & \bestfmtvalue{67}{41} & \fmtvalue{28}{31} & \fmtvalue{1}{1} & \fmtvalue{21}{29} & \fmtvalue{45}{42} \\
&SimCLR & \fmtvalue{28}{33} & \fmtvalue{1}{1} & \fmtvalue{22}{33} & \fmtvalue{44}{43} & \fmtvalue{32}{34} & \fmtvalue{1}{1} & \fmtvalue{25}{33} & \fmtvalue{49}{43} & \fmtvalue{24}{31} & \fmtvalue{1}{1} & \fmtvalue{18}{29} & \fmtvalue{38}{42} \\
&BYOL & \fmtvalue{20}{28} & \fmtvalue{1}{1} & \fmtvalue{15}{26} & \fmtvalue{36}{41} & \fmtvalue{24}{29} & \fmtvalue{1}{1} & \fmtvalue{18}{27} & \fmtvalue{42}{42} & \fmtvalue{13}{21} & \fmtvalue{1}{1} & \bestfmtvalue{8}{18} & \fmtvalue{25}{36} \\
&Barlow Twins & \fmtvalue{35}{37} & \fmtvalue{1}{1} & \fmtvalue{29}{37} & \fmtvalue{50}{44} & \fmtvalue{40}{38} & \bestfmtvalue{2}{2} & \fmtvalue{33}{39} & \fmtvalue{54}{44} & \fmtvalue{34}{37} & \fmtvalue{1}{1} & \fmtvalue{28}{37} & \fmtvalue{48}{44} \\
&VICReg & \fmtvalue{37}{38} & \fmtvalue{1}{2} & \fmtvalue{31}{39} & \fmtvalue{51}{44} & \fmtvalue{41}{38} & \bestfmtvalue{2}{2} & \fmtvalue{35}{39} & \fmtvalue{55}{44} & \fmtvalue{35}{37} & \fmtvalue{1}{1} & \fmtvalue{29}{37} & \fmtvalue{49}{44} \\
&DINO & \fmtvalue{30}{35} & \fmtvalue{1}{1} & \fmtvalue{25}{35} & \fmtvalue{46}{44} & \fmtvalue{39}{36} & \bestfmtvalue{2}{1} & \fmtvalue{32}{37} & \fmtvalue{55}{43} & \fmtvalue{27}{32} & \fmtvalue{1}{1} & \fmtvalue{21}{31} & \fmtvalue{43}{43} \\
%LeJEPA & \fmtvalue{32}{36} & \fmtvalue{1}{1} & \fmtvalue{26}{36} & \fmtvalue{47}{44} & \fmtvalue{36}{36} & \fmtvalue{1}{1} & \fmtvalue{29}{36} & \fmtvalue{52}{43} & \fmtvalue{24}{32} & \fmtvalue{1}{1} & \fmtvalue{19}{31} & \fmtvalue{37}{42} \\
&Hypersolid & \bestfmtvalue{49}{41} & \bestfmtvalue{2}{2} & \bestfmtvalue{43}{42} & \bestfmtvalue{62}{43} & \bestfmtvalue{55}{41} & \bestfmtvalue{2}{2} & \bestfmtvalue{49}{43} & \fmtvalue{66}{42} & \bestfmtvalue{51}{41} & \bestfmtvalue{2}{2} & \fmtvalue{45}{42} & \bestfmtvalue{62}{43}